\input epsf

\documentstyle[aps,floats,prl,eqsecnum,psfig]{revtex}
\tighten 

\newlength{\xyy}
\newlength{\xxyy}
\begin{document}
\draft

\twocolumn[\hsize\textwidth\columnwidth\hsize\csname @twocolumnfalse\endcsname
\preprint{AEI-2000-015, gr-qc/0003027}

\title{Gravitational waves from black hole collisions via an eclectic approach}

\author{John Baker, Bernd Br\"ugmann, Manuela Campanelli and Carlos O. Lousto}

\address{Albert-Einstein-Institut,
Max-Planck-Institut f{\"u}r Gravitationsphysik,
Am M\"uhlenberg 1, D-14476 Golm, Germany}

\date{\today}

\maketitle
\begin{abstract}
  We present the first results in a new program intended to make the
  best use of all available technologies to provide an effective
  understanding of waves from inspiralling black hole binaries in time
  for imminent observations. In particular, we address the problem of
  combining the close-limit approximation describing ringing black
  holes and full numerical relativity, required for essentially
  nonlinear interactions.  We demonstrate the effectiveness of our
  approach using general methods for a model problem, the head-on
  collision of black holes. Our method allows a more direct physical
  understanding of these collisions indicating clearly when non-linear
  methods are important.  The success of this method supports our
  expectation that this unified approach will be able to provide
  astrophysically relevant results for black hole binaries in time to
  assist gravitational wave observations.
\end{abstract}


\vskip2pc]

%\narrowtext
%\preprint{} 

Binary black hole systems pose one of the most exciting and challenging 
problems of general relativity, constituting not only a fundamental 
gravitational two-body problem, but also an important astrophysical 
problem of direct and immediate observational relevance. 
Gravitational waves from binary black hole mergers are considered one 
of the most promising candidates for experimental detection by the 
first wave of large interferometric gravitational wave observatories 
coming on line over the next few years.  These imminent observations 
present an urgent call to the theoretical relativity community to 
immediately provide any information possible about the radiation that 
might be expected from these collisions.

The problem divides physically into three phases. Initially, a slow 
adiabatic inspiral lasting nearly until the black holes are so close 
the orbital motion destabilizes, a brief period of strong essentially 
non-linear two-body interaction, and the linear ring-down of the newly 
formed remnant black hole to stationarity.  Correspondingly, 
theorists have approached the problem along three primary avenues: the 
post-Newtonian (PN) slow-motion approximation applicable in the inspiral 
phase, the 'close-limit' (CL) single perturbed black hole approximation 
handling the ring-down, and the full numerical simulation (FN) of Einstein's 
equations, which could ideally handle the entire problem on a 
large computer, but is so-far limited to brief evolutions on small 
3D domains.  Nevertheless the full numerical approach should be vital to 
treating the intermediate, essentially non-linear phase. 

In order to form the best theoretical model possible for the radiation
from these systems we feel it is vital to combine these three
approaches focusing the numerical simulations squarely on the
intermediate phase of the interaction where no perturbative approach
is applicable.  The state-of-the-art in these three fields has
advanced to the point where we can expect such an eclectic approach to
provide a reasonable model for binary black hole radiation without
depending on further advancements.  In the cases where it has been
applied the close-limit model has proven to be a reliable model for
radiation after the system has formed a common event horizon, and work
in this field has advanced sufficiently so that arbitrary
perturbations can now be calculated routinely \cite{Pullin:1999rg}.
In full numerical relativity, parts of the plunge of rather general
black hole systems, the grazing collision of two black holes with
linear momenta and spins, can be simulated\cite{Bruegmann97}.  And the
post-Newtonian method has advanced to the point where it might be
trusted even for black holes approaching the last stable orbit (LSO).
Recent estimates\cite{Buonanno00a} suggest that in the absence of
spins there are 0.6 orbits left for full-numerical treatment, and this
part of the plunge (roughly $50M$ evolution time), numerical
relativity should be able to handle today. The primary obstacle to
proceeding with the combined model is the construction of appropriate
interfaces between the three existing models.  A recent interest
within the post-Newtonian and gravitational wave research community in
providing Cauchy data for simulations may soon solve the problem of
the PN-FN interface\cite{Buonanno00a,Alvi99}.  In this letter we
introduce a general approach to providing the FN-CL interface.

The nominal result of a numerical simulation of Einstein's equations
is a time succession of values on a 3D grid for the spatial metric and
extrinsic curvature holding the geometric spacetime information. For
binary black hole simulations we expect the late time behavior of the
system to be best characterized as a ``ringing'' black hole with
outgoing radiation, with perturbation theory providing a good model
for the dynamics. The perturbative model not only allows an
inexpensive continuation of the evolution, but also supplies a clear
interpretation of the dynamics not manifest in the generic
numerical simulation. The dynamics reduces to the evolution of a
single complex field, the Newman-Penrose Weyl scalar
$\psi_4=C_{\alpha\beta\gamma\delta}
n^{\alpha}\bar{m}^{\beta}n^{\gamma}\bar{m}^{\delta}$, obeying a linear
hyperbolic equation.  Because of the axisymmetry of the background
Kerr black hole the problem can be further simplified by Fourier
decomposition of $\psi_4$, reducing to a series of 2D evolution
problems for the axial mode components of $\psi_4$ evolving
according to the Teukolsky equation\cite{Teukolsky73}.

Several important steps are required to concretely implement the 
FN-CL interface: 
1) Specify the background black hole by its mass $M$ and angular momentum 
$a=J/M^2$. 
2) Construct a space-like slice from the late-time region of the numerical 
spacetime which will be mapped to a constant time slice in the perturbative 
calculation.  In general this slice may not be related to the numerical
foliation.
3) Specify an embedding explicitly mapping the numerical slice to 
the corresponding slice in the background spacetime.  
4) Specify a (null and complex) tetrad, ($l^\mu, n^\mu, m^\mu,\bar{m}^\mu$), 
on the numerical slice which will map, on the background slice, to an 
approximation of the standard tetrad used in the perturbative formalism
where $l^\mu$ and $n^\mu$ are conveniently chosen to lie along the 
two-degenerated principal null directions of the background spacetime. 
5) Using the specified tetrad and the numerical data for the metric 
$g_{ij}$ and the extrinsic curvature $K_{ij}$ on the slice calculate 
$\psi_4$ and $\partial_t\psi_4$.  These will provide the Cauchy data 
for the perturbative evolution.  6) Evolve with the time-domain 
Teukolsky equation to determine the subsequent perturbative dynamics.
The heart of the problem is making the specifications required in (2-4).
The general idea is to numerically compute physical quantities or
geometrical invariants and relate them to their analytic expressions
in the perturbatively preferred coordinate system.  Coordinate
information can also be computed, for instance, dragging information
along geometrically defined trajectories (like geodesics) from less
problematic `faraway' parts of the spacetime where the two slices are
much closer to each other. Another possibility is to use the local
null structure of the spacetime to determine the eigenvalues of the
Weyl tensor, perturbatively related to the principal null directions
of the Kerr background.  There is generally no unique way to make
these specifications but the first order gauge and tetrad invariance
of the perturbative formalism implies that the results will not depend
strongly on small variations in these choices.  Step (5) was
explicitly worked out in Ref. \cite{Campanelli98c}.

Since a concrete implementation requires us to make choices for which
there is no clear mathematical preference, we will proceed by trying
first the simplest possible specifications and adding sophistication
only when it seems to be necessary.  We begin with a model binary
black hole problem which has already been solved by 2D numerical
relativity and close-limit perturbation theory, head-on collisions of
initially resting equal-mass black holes (Misner initial data).  At
the same time, we will try not to tune our techniques too closely to
this particular example so that our method can be readily generalized.
For this reason we will perform our numerical evolutions generically
in 3D, using well-tested, numerical techniques and codes 
(Cactus\cite{seidel98c}) ``off the shelf'' with no fine-tuning for
this problem.  We also apply perturbation theory as described by the
Teukolsky equation, allowing for a rotating black hole background,
without multipole decomposition. Specifically, for the numerical
evolutions we have used the ADM system of Einstein's equations with
maximal slicing for the lapse and vanishing shift, finite differenced
on a $128^3$ (octant mode) numerical grid, initially mapped
non-uniformly to the standard Misner coordinates to allow a distant
outer boundary without sacrificing resolution in the inner region. We
express the Teukolsky equation in Boyer-Lindquist coordinates,
although it may be convenient in the future to evolve the
perturbations in another gauge, such as a Kerr-Schild representation
of the Kerr metric \cite{Campanelli2000a}.

We implement the steps listed above as follows: 
1) In this case there is no angular momentum so the background reduces 
to Schwarzschild, $a=0$.  
Since only about 0.1\% of the system's mass will be lost  as radiation  
we specify the background mass as equal to the initial ADM mass.  
2) We make the simplest choice of background slice by identifying the 
numerical slice with a Schwarzschild time slice.  Numerical experience 
with Schwarzschild black hole evolutions in maximal slicing suggest 
a strong correspondence. 
3) For the embedding, it is clear that the trivial choice, 
identification of numerical and background coordinates is inadequate
because the black hole horizon must invariably expand in this numerical 
gauge.  On the other hand the same expansion has the tendency to drive 
the exterior region toward manifest spherical symmetry.  
A reasonable estimate for the map into the background Schwarzschild 
slice is a trivial identification of the numerical and background 
$\theta$ and $\phi$ coordinates, with some relabeling of the 
constant-$r$ spheres.  We account for the radial rescaling by 
choosing the background radius $r'$ so that the value 
of Weyl-curvature invariant  ${\cal I}=C_{abcd}C^{abcd}$ averaged 
over $\theta$ in the numerical slice coincides with 
its background value ${\cal I}=3M^2/r'^6$ in the background slice.
4) We define an appropriate, manifestly orthonormal, tetrad primarily 
by identifying timelike normal, radial, and azimuthal directions.   
The unit normal and radial direction vectors providing the spatial 
components of $\l^\mu$ and $n^\mu$. The complex vectors $m^\mu$ 
and $\bar{m}^\mu$ point within the spherical 2-surface.  At each 
step, a Gram-Schmidt procedure is first used to ensure that the 
triad remains orthonormal. Then a type III (boost) null rotation 
fixes the relative normalization of the two real-valued vectors 
to make it consistent with the tetrad assumed in the perturbative 
calculation. 5) Within the full numerical simulation we 
compute the values $\psi_4$ and $\partial_t\psi_4$ consistent 
with our tetrad specification using the formulas in 
Ref. \cite{Campanelli98c} and interpolate these Cauchy data
(using splines) to generate data directly usable by the Teukolsky 
code developed in Ref.\cite{Krivan97a}.  
For the perturbative evolutions we use $-20<r^*/M<50$ with 
$n_\theta\times n_{r^*}=48\times700$.

\begin{figure}[t]
\epsfysize=2.4in \epsfbox{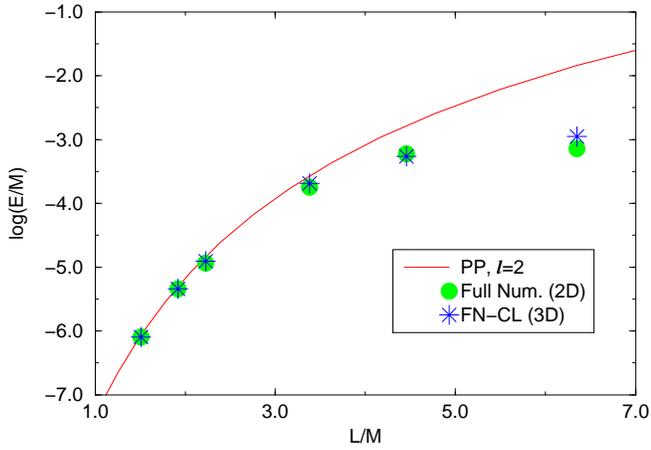}
\caption[Total radiated energy]
{ The total radiated energy from two black holes initially at rest
  (Misner initial data).  The solid line represents Price-Pullin
  results\protect\cite{Price94a} (labeled as PP, $\ell=2$) for the
  close limit approximation. 2D full numerical
  results\protect\cite{Anninos94b} are here given by full circles and
  the new FN-CL results are labeled with stars. The minimal amount of
  evolution needed for $\mu_0=1.8,2.2,2.7$ $(L\approx 3.4,4.5,6.4)$ to
  reach the perturbative regime are approximately $1M,3M,6M$
  respectively.  }
\label{fig:energies}
\end{figure}

We evolved Einstein's equations numerically from Misner initial data
for several different initial separations labeled by the parameter $\mu_0$
(corresponding to proper separations $L$ as shown in 
Fig.\ \ref{fig:energies}).
A typical duration of the total evolution was $t=10M$ and we
extracted Cauchy data every $t=1M$. A transition time $t_T$ was
determined by methods detailed below.
After each Teukolsky code evolution we  
extract the full relevant signal of the waveforms, which
typically lasted for $t<100M$.
The resulting radiation energies are shown in Fig.\ \ref{fig:energies} where
we compare  our 3D results with the results  of Ref.\ \cite{Anninos94b}
where explicit use of the symmetries of the problem have been implemented
in a 2D simulation. The other case for comparison is the
Price-Pullin\cite{Price94a} curve providing the pure close-limit result.
While all three predictions agree very well for small initial proper
separations $L/M<3$, it is clear that for larger separations the close
limit and full numerical curves deviate considerably. Our results
follow quite precisely the $2D$ computations.  A minimal full
numerical evolution time (given by our linearization time below) is
essential in obtaining the above agreement. Evolution of exact initial
data only perturbatively does not reproduce the full numerical results
for large separations, but follows the PP curve\cite{Lousto99a}.

Extracting waveforms every $1M$ of non-linear numerical evolution
allows us to study the transition to linear dynamics, and to perform
important consistency tests on our results.  If we have made a good
definition of the perturbative background in steps (1-4) above then we
can expect our radiation waveform results to be independent of the
transition time, $t_T$, once the linear regime is reached and for as
long as the numerical simulation continues to be accurate.  We apply
two independent criteria for estimating the onset of linear dynamics,
the speciality invariant prediction based only on the Cauchy data and
another estimate based on the stability of the radiation waveform
phase.  The speciality invariant introduced in \cite{Baker2000a}
predicts linear dynamics when ${\cal S}=27{\cal J}^2/{\cal I}^3$ 
differs from its
background value of unity by less than a factor of two outside the
(background) horizon.  Such a deviation from algebraic speciality
implies significant ``second order'' perturbations.  The phase of the
radiation also provides an indicator of linear dynamics.  Starting
with detached black holes, we expect an initial period of weak
bremsstrahlung radiation followed by the appearance of quasinormal
ringing. On the other hand, switching to perturbative evolution
prematurely leads to immediate ringing. Hence we first observe a
series of phase delays for the beginning of the ringing until the
actual ringing takes place, thereafter no phase shift should be seen.
The value of $t_T$ when the phase freezes gives a precise estimate of
time for linearization of the system.  We find that both estimates for
linearization time are in good agreement, yielding that $t<1M$ for the
$\mu_0=1.2$ case, $t\approx 1M$ for $\mu_0=1.8$, $3M$ for $\mu_0=2.2$,
and $6M$ for $\mu_0=2.7$.  The linearization time is somewhat longer
than the ``ringing times'' reported in Fig.\ 7 of Ref.\ 
\cite{Anninos94b} indicating that linearization occurs slightly after
the onset of ``ringing'' for the stronger collisions.  Our
linearization times are still much shorter than those for the
appearance of a common apparent horizon.
% The occurrence of a common
%{\it event} horizon should be closer to our estimates of the
%linearization time.

\begin{figure}[t]
\begin{center}
\begin{tabular}{@{}lr@{}}
\psfig{file=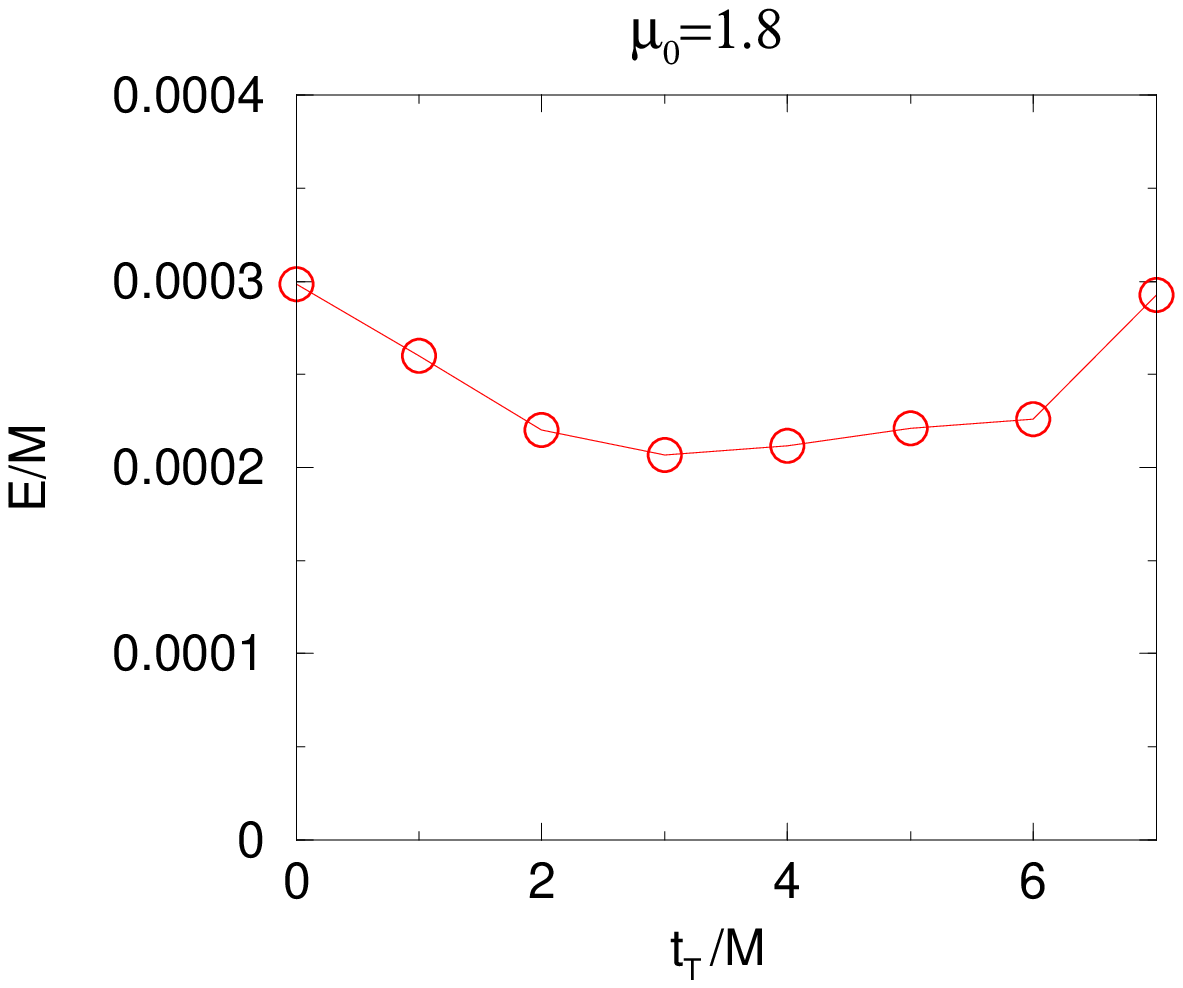,width=\xyy,clip=}
\psfig{file=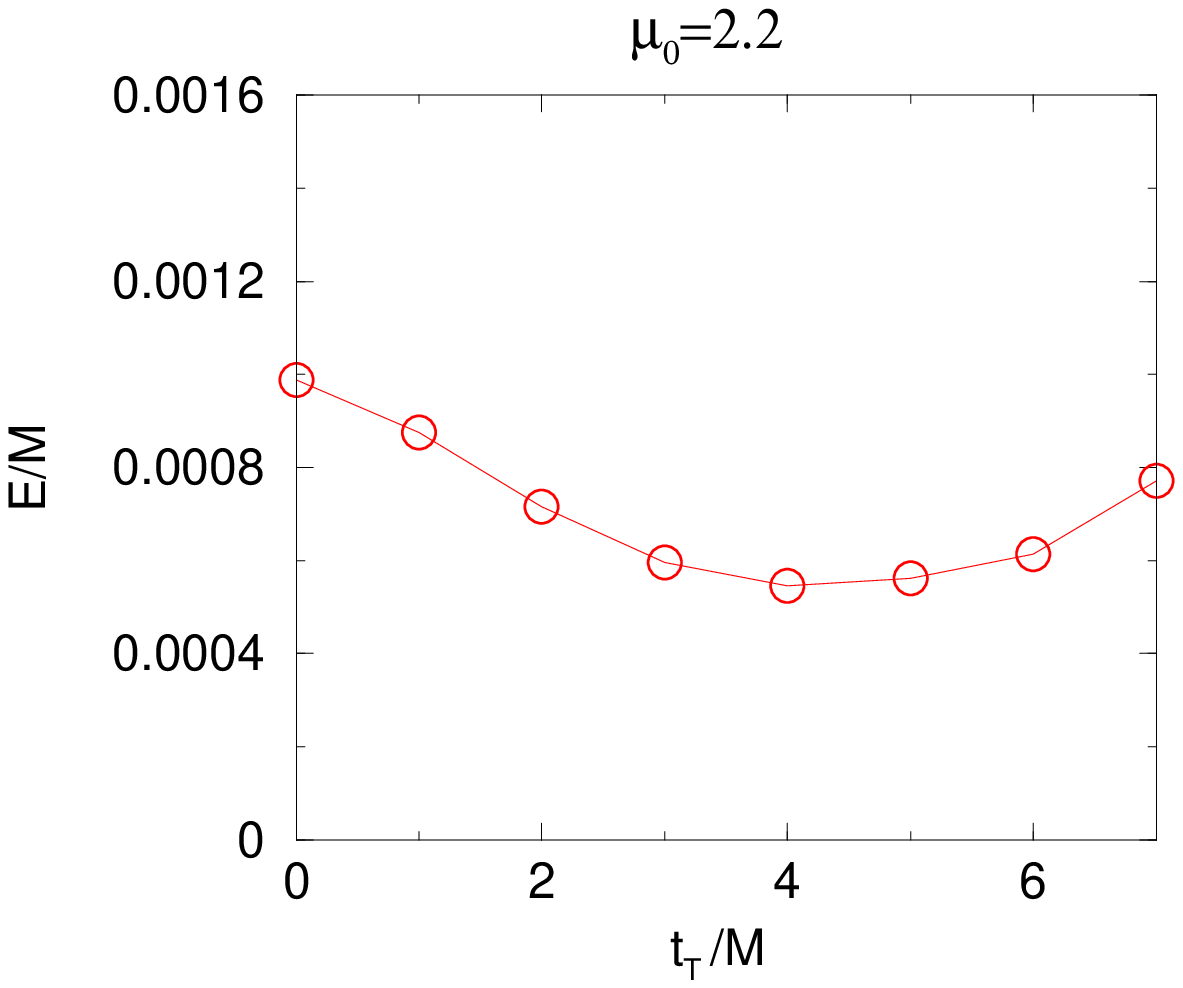,width=\xyy,clip=} 
\end{tabular}
\end{center}
\caption[]
{Radiated energy versus transition time. These figure show a clear
  plateau after linearization until numerical error begins to cause
  problems after $6M$.  }
\label{fig:evt}
\end{figure}

\begin{figure}
\begin{center}
\begin{tabular}{@{}lr@{}}
\psfig{file=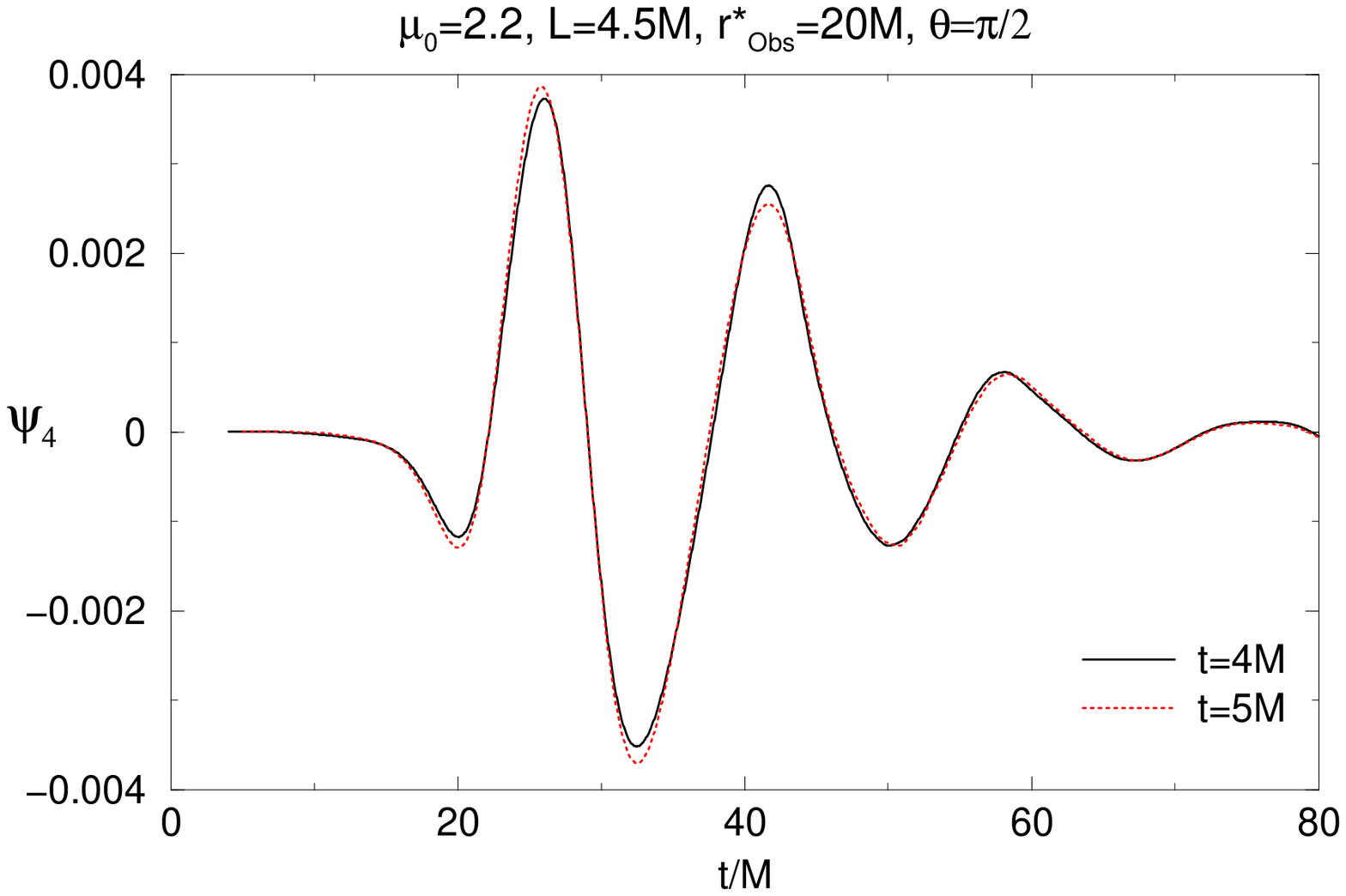,width=\xxyy,clip=}\\
\psfig{file=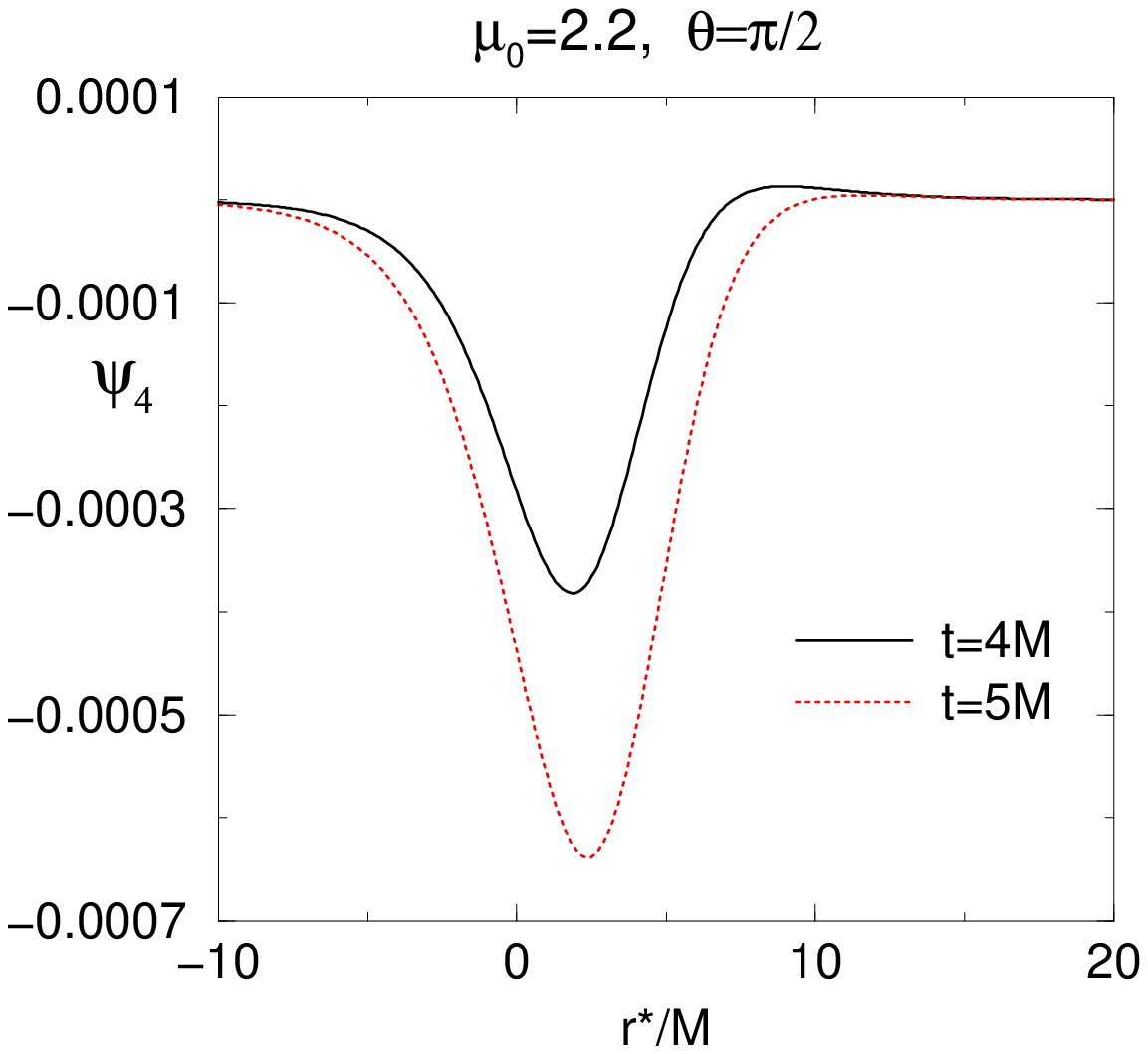,width=\xyy,clip=} 
\psfig{file=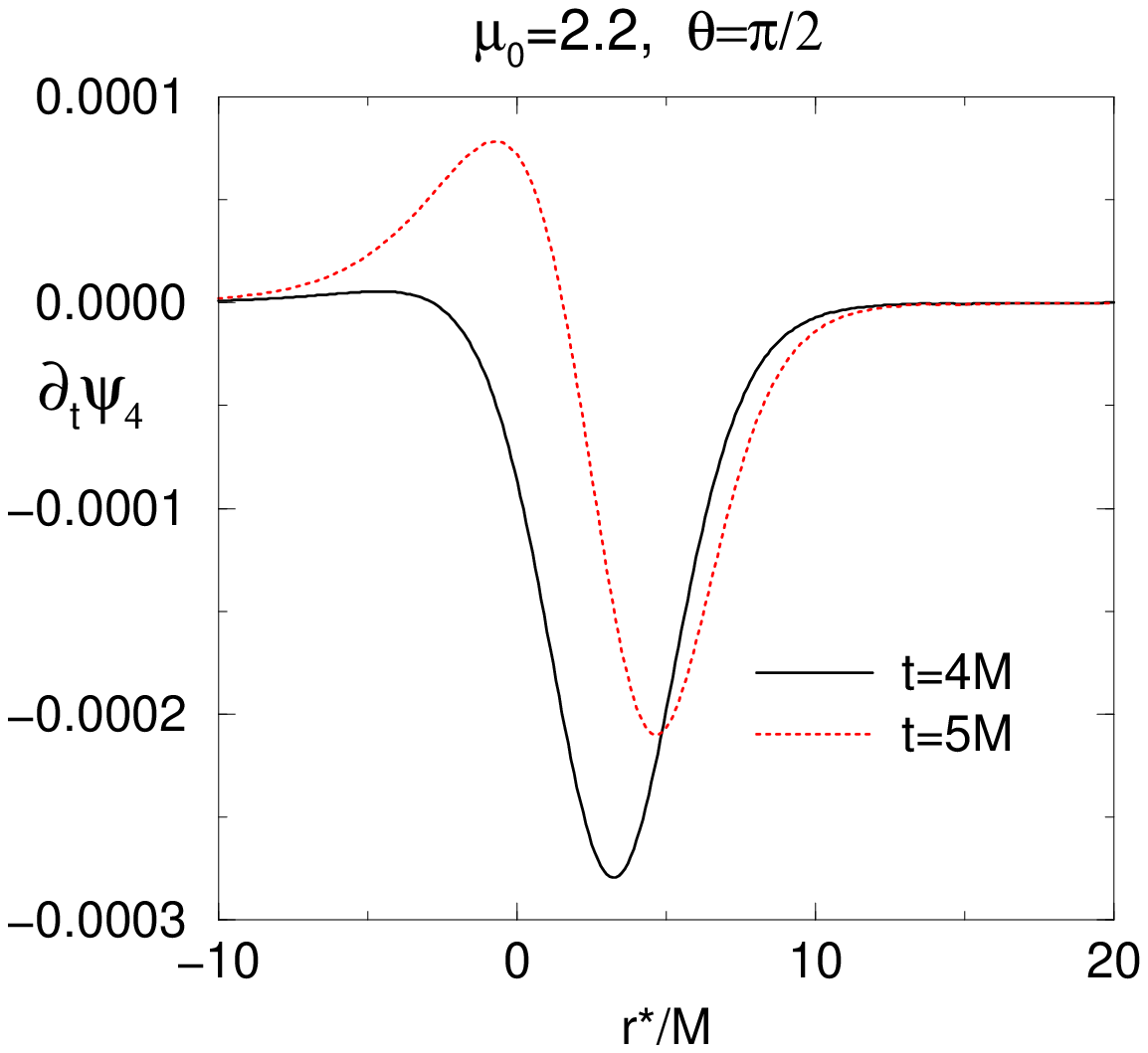,width=\xyy,clip=}
\end{tabular}
\end{center}
\caption[Waveforms]
{Waveforms and Cauchy data for initial separation $\mu_0=2.2$. 
The observer location is $r*=20M$ and $\theta=\pi/2$. The data have 
been extracted after $t=4M$ and $5M$ of full 3D nonlinear evolution 
respectively.  There are evident differences in the extracted 
Cauchy data at the two times because of the intervening evolution. 
Nevertheless the resulting waveforms agree, demonstrating the 
equivalence of the linear and non-linear evolution through this period, 
as well as the robustness of our methods. }
\label{fig:waveforms}
\end{figure}

Two example curves of energy versus $t_T$ are shown in Fig.
\ref{fig:evt}.  Before the linearization time the premature
application of perturbation theory tends to result in an overestimate
of the energy.  After linearization there is a plateau region when the
energy is insensitive to $t_T$ as is required if we have defined a
useful FN-CL interface.  Eventually, after 6M in these cases,
numerical errors caused by the ``grid-stretching'' inherent in the use
of maximal slicing tend to result again in an overestimate of the
energy.  A stronger indication of the robustness of our method is
evident in the waveforms themselves.  After linearization, the
waveforms should also be independent of $t_T$.  Fig.
\ref{fig:waveforms} shows an example of this comparing $t_T=4M$ and
$5M$ for the $\mu_0=2.2$ case.  Despite the fact that the Cauchy data
at transition time is very different, the waveform is almost
identical.  The waveform quite agrees (apart from the reversed sign)
with the $\psi_4$ published in Ref.\ \cite{Anninos94b}, Fig.\ 13. It
is worth noting here that our waveforms for the Misner data seem to be
the first complete ones computed using 3D full numerical relativity.

Perturbation theory is very useful to gain information about
waveforms from numerical spacetimes.  Customarily this interface has
been implemented only on a time-like surface to determine radiation
content ``far-away'' from the black hole.  A much more natural
boundary between the linear and non-linear regimes occurs on a
spacelike interface defined by the time beyond which non-linear black
hole perturbations no longer contribute significantly to the
radiation.  We have taken a general approach to the problem of
providing such a FN-CL interface which we believe is essential to
providing timely estimates of binary black hole waveforms. We are
aware of only one previous attempt to make a combined use of numerical
and close-limit evolution implemented in the case of two black holes
formed by collapsing matter \cite{Abrahams95d}, using a 2D
code and $l=2$ metric perturbations ({\it \`a la} Zerilli) of the
Schwarzschild background.  Our method aims toward complete generality
using full 3D numerical simulations and applying perturbation theory
as described by the Teukolsky equation, applicable to arbitrary
remnant black hole backgrounds. This approach is directly applicable
to a unified eclectic model of colliding black holes joining the
close-limit, full numerical relativity, and post-Newtonian methods.
To our knowledge this is the first time such an approach is proposed
and turned into a concrete and generic scheme.  
The success in this test case
encourages our hopes for a providing theoretical results on black hole merger 
waveforms in time to assist the first gravitational wave observations. 
We will direct our
future work toward a fully combined PN-FN-CL model for estimating
astrophysically relevant binary black hole collision waveforms.

%\acknowledgements
We wish to thank Miguel Alcubierre, Daniel Holz, Marc Mars, Richard
Price, Jorge Pullin, Bernard Schutz, Ed Seidel, and Ryoji Takahashi
for their support and many helpful discussions. M.C. was partially
supported by a Marie-Curie Fellowship (HPMF-CT-1999-00334).  All our
numerical computations have been performed on a 8 GB 32 processor SGI
Origin 2000 at AEI.

\bibliographystyle{bibtex/prsty}
\bibliography{bibtex/references}

\begin{thebibliography}{10}

\bibitem{Pullin:1999rg}
J. Pullin,   (1999), Proceedings of the Yukawa International
Symposium at Kyoto, Japan, and references therein.

\bibitem{Bruegmann97}
B. Br{\"u}gmann, Int. J. Mod. Phys. D {\bf 8},  85  (1999).
E. Seidel (1999), Proceedings of the Yukawa International
Symposium at Kyoto, Japan, and references therein.

\bibitem{Buonanno00a}
A. Buonanno and T. Damour,   (2000), gr-qc/0001013.

\bibitem{Alvi99}
K. Alvi,   (1999), gr-qc/9912113.

\bibitem{Teukolsky73}
S.~A. Teukolsky, Astrophys. J. {\bf 185},  635  (1973).

\bibitem{Campanelli98c}
M. Campanelli {\it et~al.}, Phys. Rev. D {\bf 58},  084019  (1998).

\bibitem{seidel98c}
E. Seidel and W.-M. Suen, J. Comp. Appl. Math.  (1999), in press.

\bibitem{Campanelli2000a}
M. Campanelli {\it et~al.},   , in preparation.

\bibitem{Krivan97a}
W. Krivan, P. Laguna, P. Papadopoulos, and N. Andersson, Phys. Rev. D {\bf 56},
   3395  (1997).

\bibitem{Anninos94b}
P. Anninos {\it et~al.}, Phys. Rev. D {\bf 52},  2044  (1995).

\bibitem{Price94a}
R.~H. Price and J. Pullin, Phys. Rev. Lett. {\bf 72},  3297  (1994).

\bibitem{Lousto99a}
C.~O. Lousto,   (1999), gr-qc/9911109.

\bibitem{Baker2000a}
J. Baker and M. Campanelli,  (2000), gr-qc/0003031.

\bibitem{Abrahams95d}
A.~M. Abrahams, S.~L. Shapiro, and S.~A. Teukolsky, Phys. Rev. {\bf D51},  4295
   (1995).

\end{thebibliography}
\thebibliography{PRL1}

\end{document}